# Ab initio calculation of UV-Vis absorption spectra of a single molecule chlorophyll *a*: Comparison study between RHF/CIS, TDDFT, and semi-empirical methods

**[1]* Veinardi Suendo & [2] Sparisoma Viridi**

[1]Inorganic and Physical Chemistry Research Division, Department of Chemistry, Faculty of Mathematics and Natural Sciences, Institut Teknologi Bandung, Jalan Ganesha 10, Bandung 40132, Jawa Barat, Indonesia
[2] Nuclear Physics and Biophysics Research Division, Department of Physics, Faculty of Mathematics and Natural Sciences, Institut Teknologi Bandung, Jalan Ganesha 10, Bandung 40132, Jawa Barat, Indonesia
*Corresponding author. Tel.: +62 222502103, Fax: +62 222504154, email: vsuendo@chem.itb.ac.id

**Abstract.** Chlorophyll *a* is one the most abundant pigment on Earth that responsible for trapping the light energy to perform photosynthesis in green plants. This molecule is a metal-complex that consists of a porphyrin ring with high symmetry that acts as ligands with magnesium as the central ion. Chlorophyll *a* has been studied for many years from different point of views for both experimental and theoretical interests. In this study, the restricted Hartree-Fock configuration interaction single (RHF/CIS), time-dependent density functional theory (TDDFT) and some semi-empirical methods (CNDO/s and ZINDO) calculations were carried out and compared to reconstruct the UV-Vis absorption spectra of chlorophyll *a*. In some extend, the calculation results based on a single molecule calculation were succeeded to reconstruct the absorption spectra but required to be scaling to match the experimental one. Different computational methods (ab initio and semi-empirical) exhibit the differences in the energy correction factor and the presence of transition states, however, still conserve the main spectral features. In general, the semi-empirical methods provide a better energy scaling factor, which means closer to the experimental one. However, they lack of fine features or vertical transitions with respect to the experimental spectra. The ab initio calculations result more complete features than the semi-empirical methods, especially the TDDFT with high level of basis sets that provides a good accuracy in transition energies. The contribution of ground states and excited states orbitals in the main vertical transitions is discussed based on orbital structure. This might gives a new perspective to explain the energy transfer phenomena in the absorption processes that related to the function of chlorophyll a as a light-harvesting antenna.

## 1 Introduction

Chlorophyll *a* is one the most abundant pigment on Earth that responsible for trapping the light energy to perform photosynthesis. This molecule is a metal-complex that consists of a porphyrin ring that acts as ligands with magnesium as the central ion (Figure 1). Chlorophyll *a* has been studied for many years from different point of views for both experimental and theoretical interests. Its intrinsic spectroscopic, magnetic and electrochemical properties have attracted many researchers to carry out experiments, including researches on the applications of this natural pigment. On the other hand, the molecular model of this compound, porphyrins metal complex or metal porphyrin, has a structure with high molecular symmetry of $D_{4h}$, which is very interesting from the theoretical point of view. In some extend, the theoretical studies of the molecular model can explain the general features of metal porphyrins. However, the influences of the attached peripheral groups, especially that break the symmetry, such as in chlorophyll *a*, play an important role in their function. These influences can be seen clearly, how the absorption/emission spectra of chlorophyll *a* differs from chlorophyll *b*, while both of them still conserve the same main features.

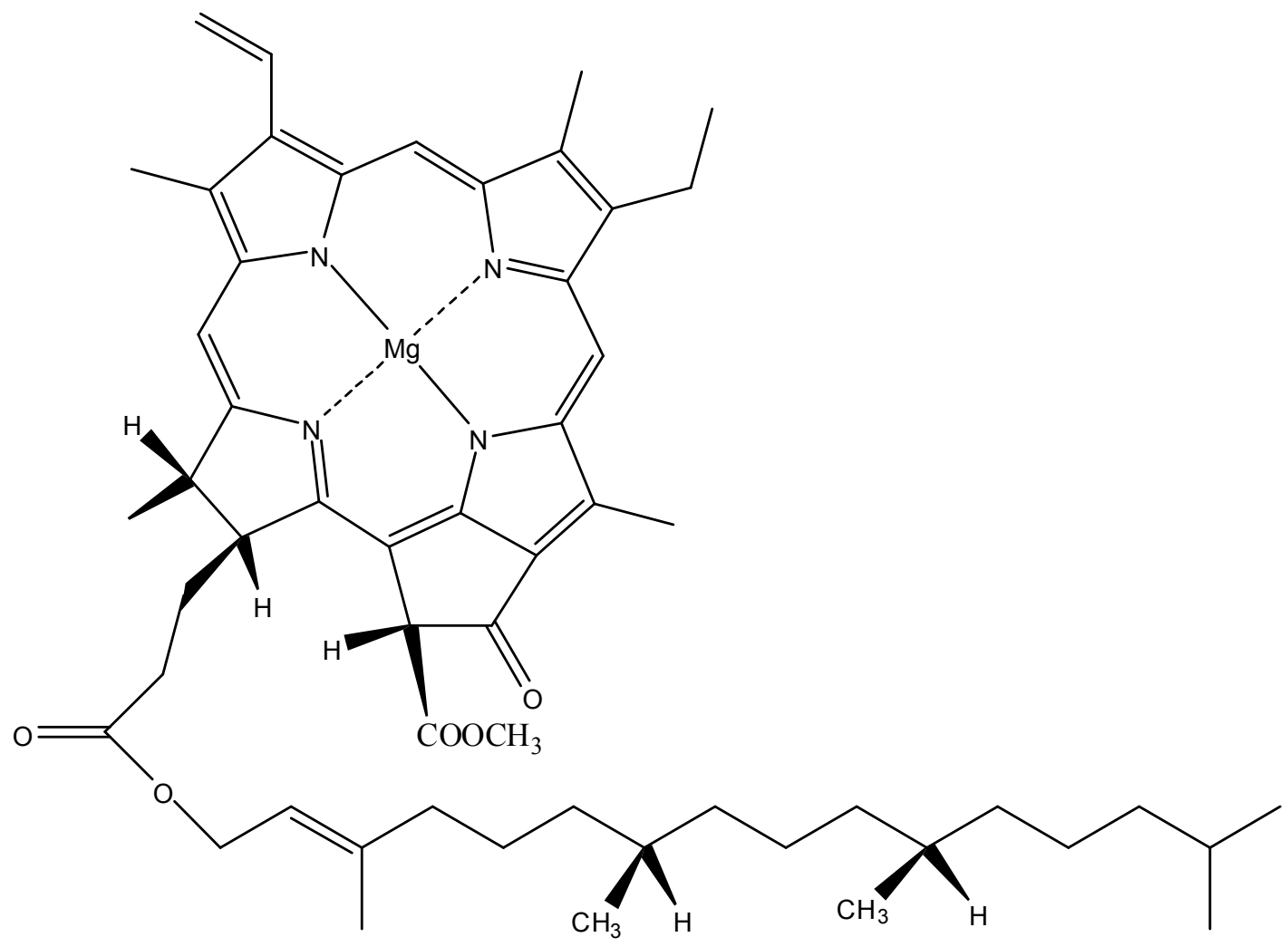

**Figure 1** Structural formula of the chlorophyll *a* molecule

The ability of this molecule to convert the light into electricity and to induce redox reactions is very interesting to be implemented in artificial systems. Thus, this molecule is a prospective candidate as a base material for the many new environmental friendly devices, such solar cell [1] or light emitting diode [2]. However, the main issues concerning the implementation of this molecule in the real life are the efficiency and stability of the molecule itself to function in the device. These properties much depend on the nature of both ground states and excited states orbitals. Thus, knowing the electronic states of this molecule will reveal a better understanding about their functions related to the electronic excitation-deexcitation processes involving the ground states and excited states orbitals at various environments, such as in solution, in photosynthetic light-harvesting protein complexes and in well-function electronic devices. As an example, it is not well understood why the $Q_x$ and Soret Bands are much more sensitive to solvent effects than the $Q_y$ band [3]. This is a real orbitals interaction problem, which has to be solved quantum mechanically and rather difficult to be studied directly through experiments. Thus, the ab initio calculation is one of the good solutions that must be performed here, to study the excited states and possible transitions.

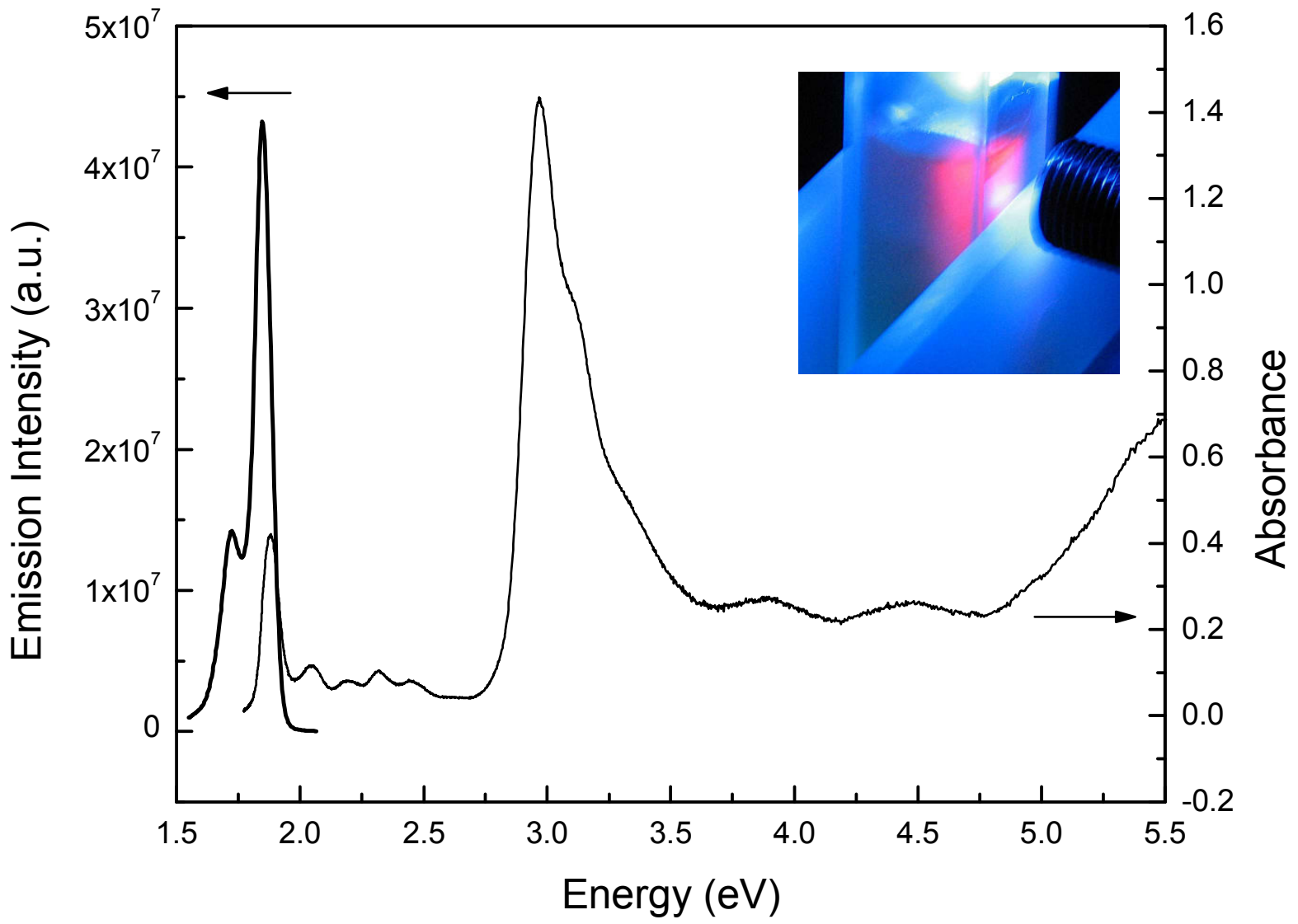

**Figure 2** Fluorescence and absorption spectra of chlorophyll *a* in methanol [4]. Image in the inset shows a strong red emission from chlorophyll *a* solution excited near the Soret band at 405 nm.

The experimental spectra of chlorophyll a molecule consists of two main absorption bands (Figure 2). The more complex one is located in the blue region, called Soret band that consists of many electronic transitions with strongest oscillator strength at 2.88 eV called *B* transition [3,5]. The other band is called *Q* band that found in red region that consists of $Q_y$ and $Q_x$ transitions at 1.87 and 2.14 eV, respectively [3,5]. The electronic spectra calculation of chlorophyll *a* have been studied intensively with high accuracy [6] and compared with its pheophytin form [7]. Here, the calculation results will hopefully lead to construct a better model to optimize the design of new materials and devices. In this study, we focused on the variation two calculation methods: the configuration interaction singles (CIS), which is well known as a low-cost calculation method; and the time-dependent density functional theory (TDDFT), which is a more reliable method. Here, the basis set that used to construct the molecular wave function is varied at three polarization levels of 6-31G, 6-31G(d) and 6-31G(d,p). As comparison some semi-empirical methods (CNDO/s and ZINDO) were also carried out to calculate the UV-Vis absorption spectra in order to have a clear perspective about the accuracy of methods under study.

## 2 Methods

### 2.1 Configuration Interaction Singles (CIS) [8,9]

The CIS calculation can be formulated by starting from HF (Hartree-Fock) ground state, $\Phi_0(\mathbf{r})$, which is corresponds to the best single Slater determinant describing the electronic ground state of the system, which can be written

$$\Phi_0(\mathbf{r}) = \left|\phi_1(\mathbf{r})\phi_2(\mathbf{r})\phi_3(\mathbf{r})\ldots\phi_n(\mathbf{r})\right| \tag{1}$$

A closed-shell ground-state electronic configuration is normally assumed to simplify the problem, where the $\phi_i(\mathbf{r})$ correspond to doubly occupied spatial orbitals, $n = N/2$ and $N$ is the number of electrons. Here, $\Phi_0(\mathbf{r})$ is obtained by solving the time-independent Hartree-Fock equation, which is given by

$$\hat{F}(\mathbf{r})\Phi_0(\mathbf{r}) = E_0\Phi_0(\mathbf{r}) \tag{2}$$

In the configuration interaction, the electronic wave function is then constructed as a linear combination of the ground state Slater determinants and so-called excited determinants, which are obtained by replacing occupied orbitals of the ground state with the virtual ones. If one replaces only one occupied orbital *i* by one virtual orbital *a* and then includes only these Slater determinants, $\Phi_i^a(\mathbf{r})$ in the CI wave function expansion, thus, we obtain the CIS wave function, $\Psi_{CIS}$, which can be written

$$\Psi_{CIS} = \sum_{ia} c_i^a \Phi_i^a(\mathbf{r}) \tag{3}$$

The CIS energy can be obtain readily by introducing the CIS wave function, $\Psi_{CIS}$ into the exact time-independent electronic Schrödinger equation,

$$\hat{H}(\mathbf{r})\Psi_{CIS}(\mathbf{r}) = E_{CIS}\Psi_{CIS}(\mathbf{r}) \tag{4}$$

Finally, one readily obtains an expression for the excitation energies,

$$\omega_{CIS} = E_{CIS} - E_0 \tag{5}$$

## 2.2 Time-Dependent Density Functional Theory [8,10,11]

The TDDFT calculation can be formulated by the time-dependent Kohn-Sham equation (TD-KS)

$$i\frac{\partial}{\partial t}\phi_i(\mathbf{r},t) = \left(-\tfrac{1}{2}\nabla_i^2 + v(\mathbf{r},t) + \int d^3\mathbf{r}\frac{\rho(\mathbf{r}',t)}{|\mathbf{r}-\mathbf{r}'|} + \frac{\delta A_{xc}[\rho]}{\delta\rho(\mathbf{r},t)}\right)\phi_i(\mathbf{r},t) = \hat{F}^{KS}\phi_i(\mathbf{r},t) \tag{6}$$

with

$$\rho(\mathbf{r},t) = \rho_S(\mathbf{r},t) = \sum_i^N |\phi_i(\mathbf{r},t)|^2 \tag{7}$$

Two different strategies can be followed to obtain excitation energies and oscillator strengths by employing the TD-KS. The first possibility is to propagate the TD-KS wave function in time, which is referred as real-time TDDFT. The second approach is the analysis of the linear response of the TD-KS equation that used in this work.

For the linear response of TD-KS equation, let us suppose there exists a time-dependent perturbing potential $v_1(\mathbf{r},t)$, for an oscillating electric field $v_1(\mathbf{r},t) = E_z \cos wt$, which is switched on at time $t = t_0$. The external potential can be expressed as

$$v_{ext}(\mathbf{r},t) = v_0(\mathbf{r}) + v_1(\mathbf{r},t) = \begin{cases} v_0(\mathbf{r}) & ,t \le t_0 \\ v_0(\mathbf{r}) + E_z \cos wt, & t > t_0 \end{cases} \quad (8)$$

where $v_0$ is the Coulomb potential between electrons and nuclei

$$v_0(\mathbf{r}) = -\sum_K^N \frac{Z_K}{|\mathbf{R}_K - \mathbf{r}|} \quad (9)$$

Thus, the first order deviation of the time-dependent density, $\rho(\mathbf{r},t)$ from the unperturbed ground state density $\rho_0(\mathbf{r})$ for interacting particles can be written as

$$\rho(\mathbf{r},t) - \rho_0(\mathbf{r}) \approx \rho_1(\mathbf{r},t) = \int dt' \int d\mathbf{r}' \chi(\mathbf{r},t,\mathbf{r}',t') v_1(\mathbf{r}',t') \quad (10)$$

with the interaction response function

$$\chi(\mathbf{r},t,\mathbf{r}',t') = \left. \frac{\delta\rho(\mathbf{r},t)}{\delta v_{ext}(\mathbf{r}',t')} \right|_{v_0} \quad (11)$$

while for non-interacting particles, we will have

$$\rho_1(\mathbf{r},t) = \int dt' \int d\mathbf{r}' \chi_s(\mathbf{r},t,\mathbf{r}',t') v_{s,1}(\mathbf{r}',t') \quad (12)$$

with the Kohn-Sham response function

$$\chi_s(\mathbf{r},t,\mathbf{r}',t') = \left. \frac{\delta\rho(\mathbf{r},t)}{\delta v_s(\mathbf{r}',t')} \right|_{v_s[\rho_0]} \quad (13)$$

Here, the poles of the response function (eq. 11) of the interacting system represent the electronic transition energies.

## 2.3 Computational Detail

PM3, RHF, RHF/CIS and TDDFT calculations of chlorophyll *a* were carried out using the PC GAMESS/Firefly program package developed by Alex A. Granovsky and co-workers at Department of Chemistry, Moscow State

University [12]. Semiempirical quantum calculations of UV-Vis absorption spectra were carried out as well using Winmostar (CNDO/S [13,14]) and Arguslab (ZINDO) [15-17] software packages. The starting structure of chlorophyll *a* in CLA_model.pdb file is obtained from RSCB Protein Data Bank (http://www.rcsb.org/pdb/home/home.do). Here, all calculations of the chlorophyll *a* were carried out using the structure in the absent of other ligands, such as $H_2O$, $NH_3$ or Cl- in the axial positions, in order to simplify the calculation that also supported from the experimental evidence by presence of this structure in nature from ESI-MS (electro-spray ionization mass spectroscopy) measurements. This structure was geometrically optimized in two steps, firstly using semiempirical method at PM3 level then followed by RHF using 6-31G(d) basis set until reach a minimum in energy prior to any further properties calculations. The electronic transitions between occupied and unoccupied state were calculated at RHF/CIS and TDDFT/B3LYP level of theory using basis set 6-31G, 6-31G(d) and 6-31(d,p) that result the UV-Vis absorption spectra and the main orbitals that contribute to the transitions.

# 3 Results and Discussion

## 3.1 Commonly observed transitions

Figure 2 shows both absorption and emission spectra of chlorophyll *a* that obtained experimentally in methanol [4]. In the absorption spectra, we can observe the presence of $B$ (Soret), $Q_y$ and $Q_x$ transitions at 2.88 eV, 1.87 and 2.14 eV, respectively [3,4]. Meanwhile, a strong emission is observed around $Q$ band or in the red region (see inset Figure 2). These features may indicate that the molecular orbitals involved in $Q$ transitions are responsible for the emission processes. Further, it is also observed that the excitation energies for fluorescence measurements have to be overlapped with the Soret band. This reveals the presence of energy transfer between states that involved in transitions of $Q$ and Soret bands. The SAAP (spin-adapted antisymmetrized product) treatment on both RHF/CIS and TDDFT calculations reveals that the main molecular orbitals contributed in the electronic transitions are HOMO-1, HOMO, LUMO and LUMO+1 independent of level of theory and basis sets.

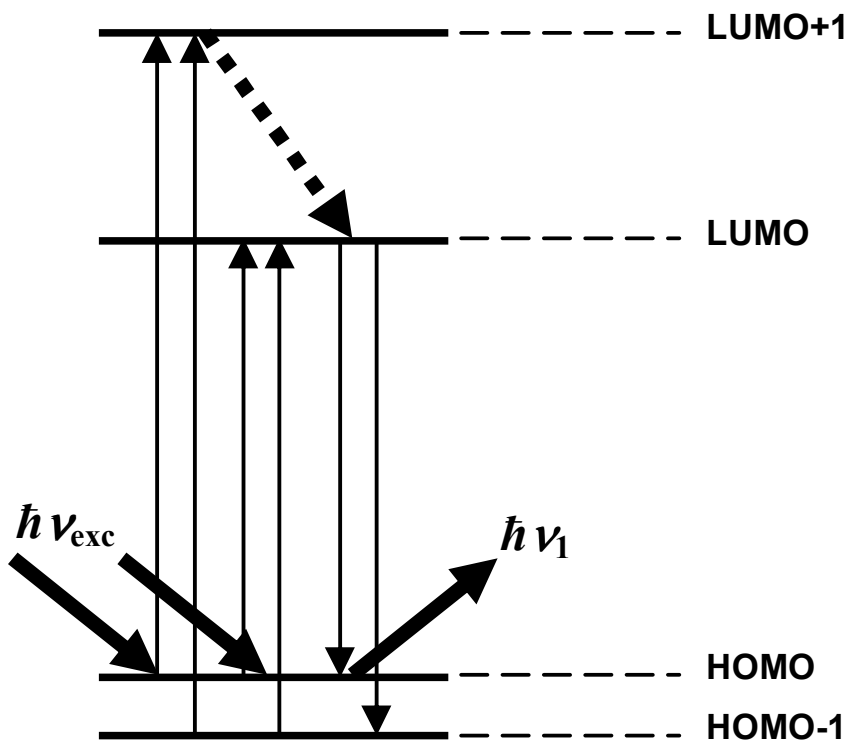

**Figure 3** Schematic diagram of excitation and de-excitation processes involving two highest occupied and two lowest unoccupied states in a chlorophyll *a* molecule.

This result is in a good agreement with the shape of observed UV-Vis absorption spectra that presented in Figure 2, which has two main absorption regions represent *Q* and Soret Bands that can be assigned for the transitions from HOMO and/or HOMO-1 to LUMO and LUMO+1, respectively. Thus, the scheme of upward vertical electronic transitions in chlorophyll *a* can be simplified into four main transitions that assigned for two absorption bands *Q* and Soret, respectively:

$$\left.\begin{array}{lcl}\text{HOMO} & \rightarrow & \text{LUMO}\\ \text{HOMO}-1 & \rightarrow & \text{LUMO}\end{array}\right\} \quad Q \text{ band}$$

$$\left.\begin{array}{lcl}\text{HOMO} & \rightarrow & \text{LUMO}+1\\ \text{HOMO}-1 & \rightarrow & \text{LUMO}+1\end{array}\right\} \quad \text{Soret Band}$$

Further, by combining the both experimental data and calculation results with the Franck-Condon principle, the simplified mechanism for downward vertical transition can also be predicted as

$$\left.\begin{array}{lcl}\text{HOMO} & \leftarrow & \text{LUMO}\\ \text{HOMO}-1 & \leftarrow & \text{LUMO}\end{array}\right\} \quad \text{Emission band}$$

Here, the emission band is located near *Q* band, but slightly red shifted due to the relaxation of excited structure as explained by Franck-Condon principle. The excited structure has a strong tendency to relax its structure, to achieve minimum in energy before their electrons deexcited to the ground states radiatively. Figure 3 shows all the simplified vertical electronic transitions of chlorophyll *a*, which cover *Q*, Soret and emission bands. The downward transition scheme is also has a good agreement with the experimental data due to the presence to two main emission features in the observed emission band at 1.72 and 1.85 eV that assigned for the relaxed form of $Q_y$ and $Q_x$ transitions, respectively (Figure 2).

Some fine features are also presented in the experimental spectra, which appear between *Q* and Soret bands. These features are not so important in some processes, i.e. absorption or emission processes, but they are very important indicator to select a good model to calculate the electronic structure, especially in the study of electronic transitions. They have the opposite behavior compared to the *Q* or Soret transitions, where these fine transitions are very sensitive to the wavefunction that used in calculations. Their presence depends much on which wavefunction used, whether it is enough or not to have orbital interactions needed for certain electronic transitions to occur. This might be explained by the coupling of the vibrational modes on the electronic transitions that highly depend on the orbital interactions. Thus, they need a model with high level of theory and large basis sets, where the polarization effect, representing the orbital interactions is not negligible. Further discussion about this topic is presented in the next section of this paper.

## 3.2 Molecular orbitals

Figure 4 and 5 show the four main orbitals that contributed in the vertical electronic transitions in chlorophyll *a*, which calculated in the basis set level of 6-31G(d,p) using RHF/CIS and TDDFT methods, respectively. These orbitals namely: HOMO-1, HOMO, LUMO and LUMO+1, which represent two highest occupied orbitals and two lowest unoccupied orbitals in chlorophyll *a* molecule. Here, we only presented the graphical representation of calculated orbitals for the calculations using the highest level of basis set, 6-31G(d,p), which result the best fit in UV-Vis spectra with experimental one. This basis set includes the polarization effects of the unoccupied orbitals *d* and *p* on the occupied orbital *p* and *s* in the molecules, respectively. Thus, the constructed molecular wavefunction has more sophisticate orbital interactions that consequently provide more complete possible electronic transitions as presented in the calculated UV-Vis absorption spectra in the next section.

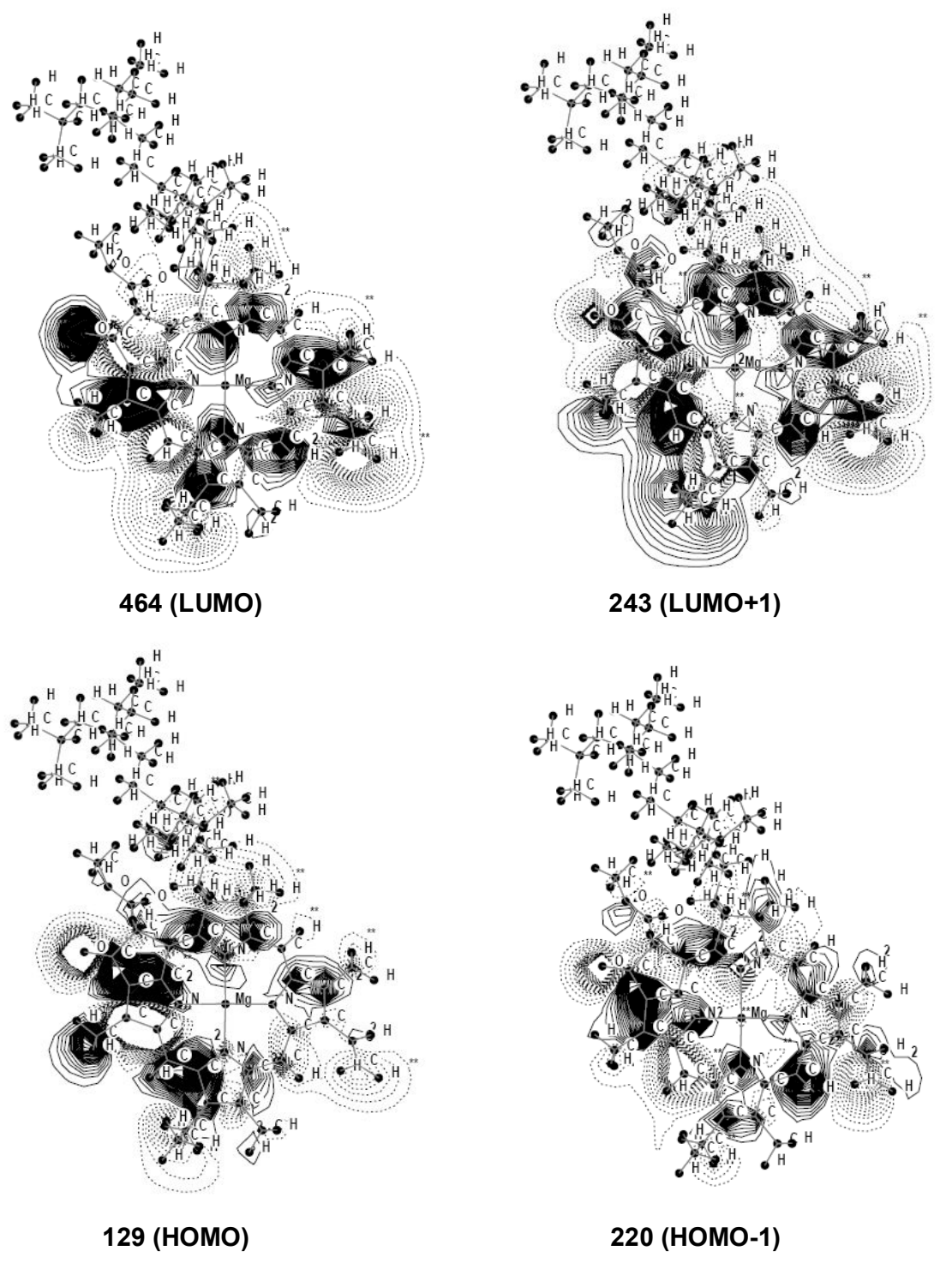

**Figure 4** Two highest occupied orbitals and two lowest unoccupied orbitals in a chlorophyll *a* molecule calculated at RHF/CIS using 6-31G(d,p) basis set.

In general, if we compare the molecular orbitals presented in Figure 4 and 5, they provide almost the same spatial distribution of electron in three-dimensional space. The both calculations provide similar results, where the frontier orbitals of the molecules are centered on the porphyrin ring but not on the phytol chain. This gives a consequence that the structure of the porphyrin ring will be responsible for most of electronic transitions involved by chlorophyll *a* molecule, such as absorption, emission and charge transfer. The similar spatial distribution of orbital lobi between the pairs of HOMO/HOMO-1 and LUMO/LUMO+1, reveals the strong contribution of structural symmetry of porphyrin ring that still exists in the chlorophyll *a*, even with the presence of functional groups such as phytol chain, methine, methyl esther, methyl, ethyl and ethylene on the periphery of the ring. The similar spatial distribution between occupied and unoccupied frontier orbitals mostly contribute in electronic transition of the molecule involving photons that provides high oscillator strengths for both upward and downward transitions.

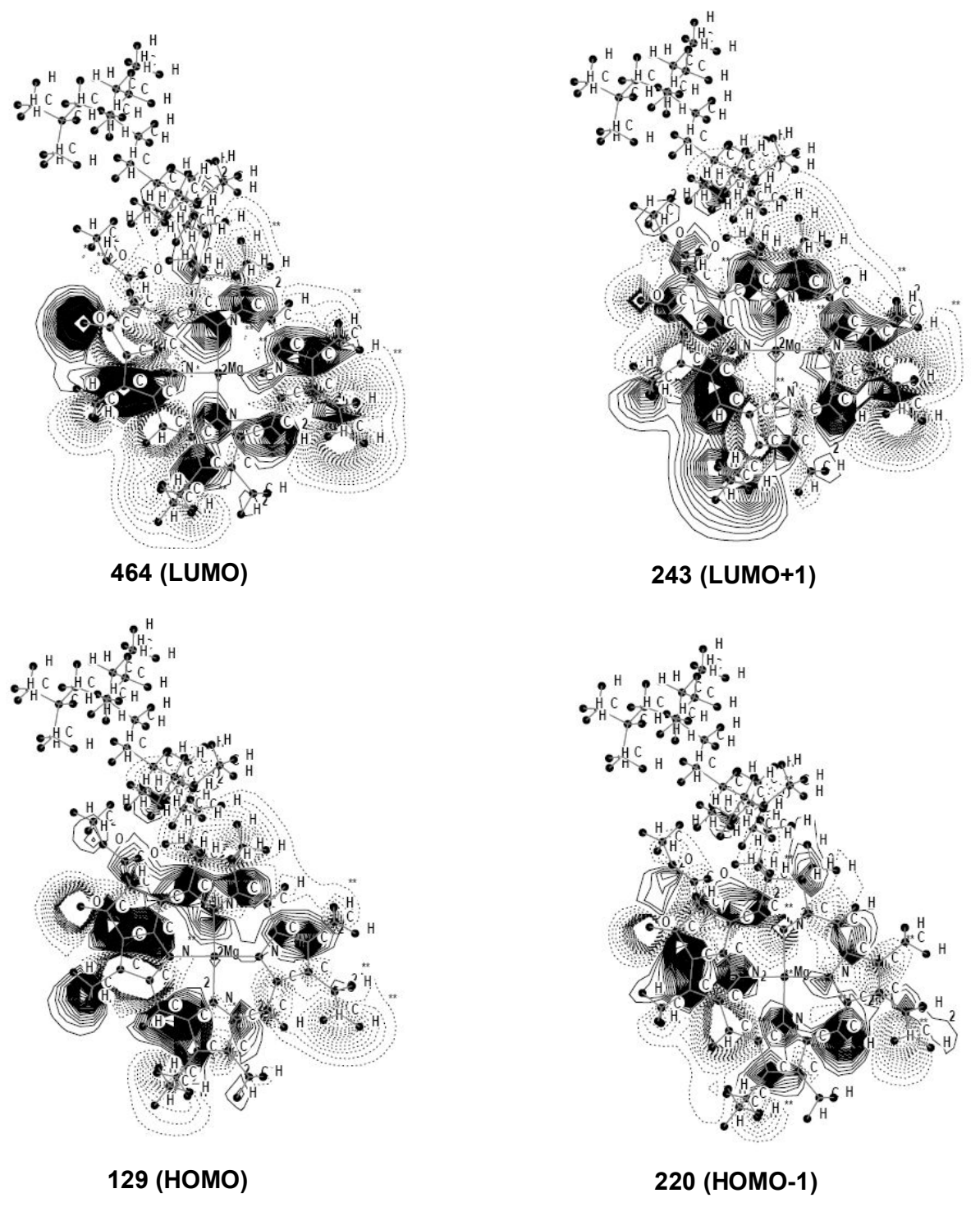

**Figure 5** Two highest occupied orbitals and two lowest unoccupied orbitals in a chlorophyll *a* molecule calculated at TDDFT using 6-31G(d,p) basis set.

Figure 6 summarized the energy levels of the main molecular orbitals that involved in the electronic vertical transitions. Here, the TDDFT method provides more accurate results than RHF/CIS with respect to experiment, where we have a strong absorption/emission in the red range (≈ 2 eV). The RHF/CIS method tends to over estimate the energy gap between HOMO and LUMO. This can be slightly improves by increasing the polarization level of the basis set, where this increase the energy of HOMO faster than LUMO, results an increase in the energy gap. The same improvement can be observed as well in the case of TDDFT calculations. In this case, the effect is more pronounce due to the energy level of LUMO that nearly independent of polarization level. Moreover, based on the MO analysis, we observed that in the basis set level of 6-31G(d,p), the order of molecular orbitals around HOMO/LUMO will be independent of the level of calculation theory (Figure 4 and 5). This suggests that the improvement by polarization level of the basis set reached saturation for this molecule. Thus, the further improvement has to be made either by increasing the level of theory or basis set level, such adding the localization or diffusion function.

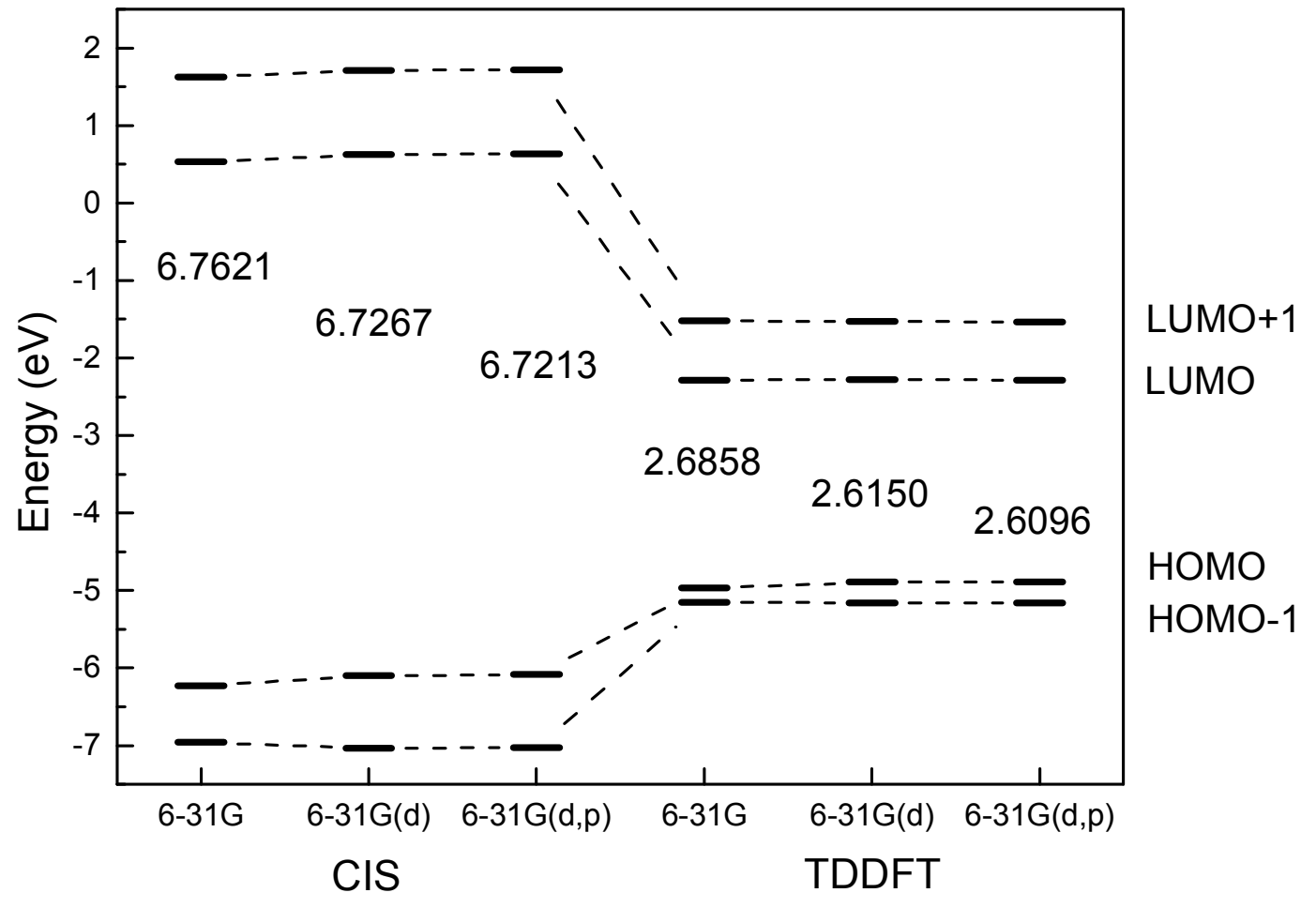

**Figure 6** Energy level of two highest occupied orbitals and two lowest unoccupied orbitals in a chlorophyll *a* molecule calculated at different level of theories and basis sets.

Figure 6 also provides the information that RHF/CIS method tends to over estimate the interactions among AOs that form the frontier orbitals without involving long interactions among formed MOs. Thus, RHF/CIS gives better stabilization on HOMO and HOMO-1, but worst on LUMO and LUMO+1, resulting an over estimating energy gap between HOMO and LUMO. In the case of TDDFT calculations, which in principle using a DFT method that has a delocalization feature in their molecular wavefunctions, thus naturally provides a better long range interactions among MOs, as presented by the less stabilized HOMO and HOMO-1, while LUMO and LUMO+1 are more stabilized, with respect to RHF/CIS results. The delocalization feature of DFT is also obvious on the effect of polarization on the unoccupied orbitals, where the addition of polarization functions give only a small contribution on the orbital stabilization. This can be explained by the nature of unoccupied orbitals that are more delocalized and spatially spread over a large region, especially in the DFT formulation. As the consequence, this will reduce the effect of polarization functions significantly. The polarization functions that act as a correction factor to add the interaction between occupied and unoccupied orbitals in their AOs form will lose their effect as the molecular wavefunctions get delocalized spatially. The delocalized wavefunctions might provide more interactions among MOs that in some extent results a better calculation of electronic transitions with respect to experiment.

### 3.3 Calculated UV-Vis absorption spectra

Figure 7 shows the calculated UV-Vis absorption spectra of chlorophyll *a* at different level of theories and basis sets, which includes two semi-empirical methods, CNDO/s and ZINDO. All calculated spectra are corrected using their maximum transition with respect to the position of *B* transition of Soret Band at 2.88 eV. The correction factor for each calculation is tabulated in Table I. The closer correction factor to unity provides a higher accuracy of the calculation method. All methods give the correction factor less than unity that indicates the over estimation on the upward vertical electronic transition energies, that represented by the UV-Vis absorption spectra. In general, the semi-empirical methods provide a closer correction factor among all methods in this study. It is a common feature due the nature of semi-empirical wavefunctions, which has been parameterized, especially CNDO/s and ZINDO, for the study of electronic transitions involving photon [13-17].

In some cases, the usage of semi-empirical methods is more than enough to calculate the UV-Vis absorption spectra, especially if it only needs to provide the spectra qualitatively that consists of the main electronic transitions. However, if we need to study the fine features of electronic transition, such as transitions that appear between *Q* and Soret bands in chlorophyll *a*, the ab initio methods are the better solution. Figure 7 compares the spectra calculated with all methods provide in this study. It is obvious that ab intio methods, especially TDDFT and high level of polarization provide more transition peaks in the region between *Q* and Soret bands. As discussed in the previous section, in principle the DFT treatment also provides more orbital interactions as the polarization functions, but in much larger region spatially. Based on its low occurrence, we predicted that the electronic transitions assigned for fine features in UV-Vis absorption spectra of chlorophyll *a* have a low probability are due to the mismatch of the orbital symmetry. This effect can be reduce if the molecular wavefunctions become more loose or in other words, the long interaction among MOs in become more pronounce or stronger. This provides a condition that will not hold strictly the symmetry conservation during electronic transition due the uncertainty of their distribution in space. Thus, both DFT treatment and the addition of polarization functions in molecular wavefunctions might improve the calculation results. Moreover, this explanation is in a good agreement with the nature of chlorophyll *a*, a molecule with delocalized electrons, which provide more long range interactions among MOs that might be the origin of the fine features in the UV-Vis absorption spectra between *Q* and Soret bands.

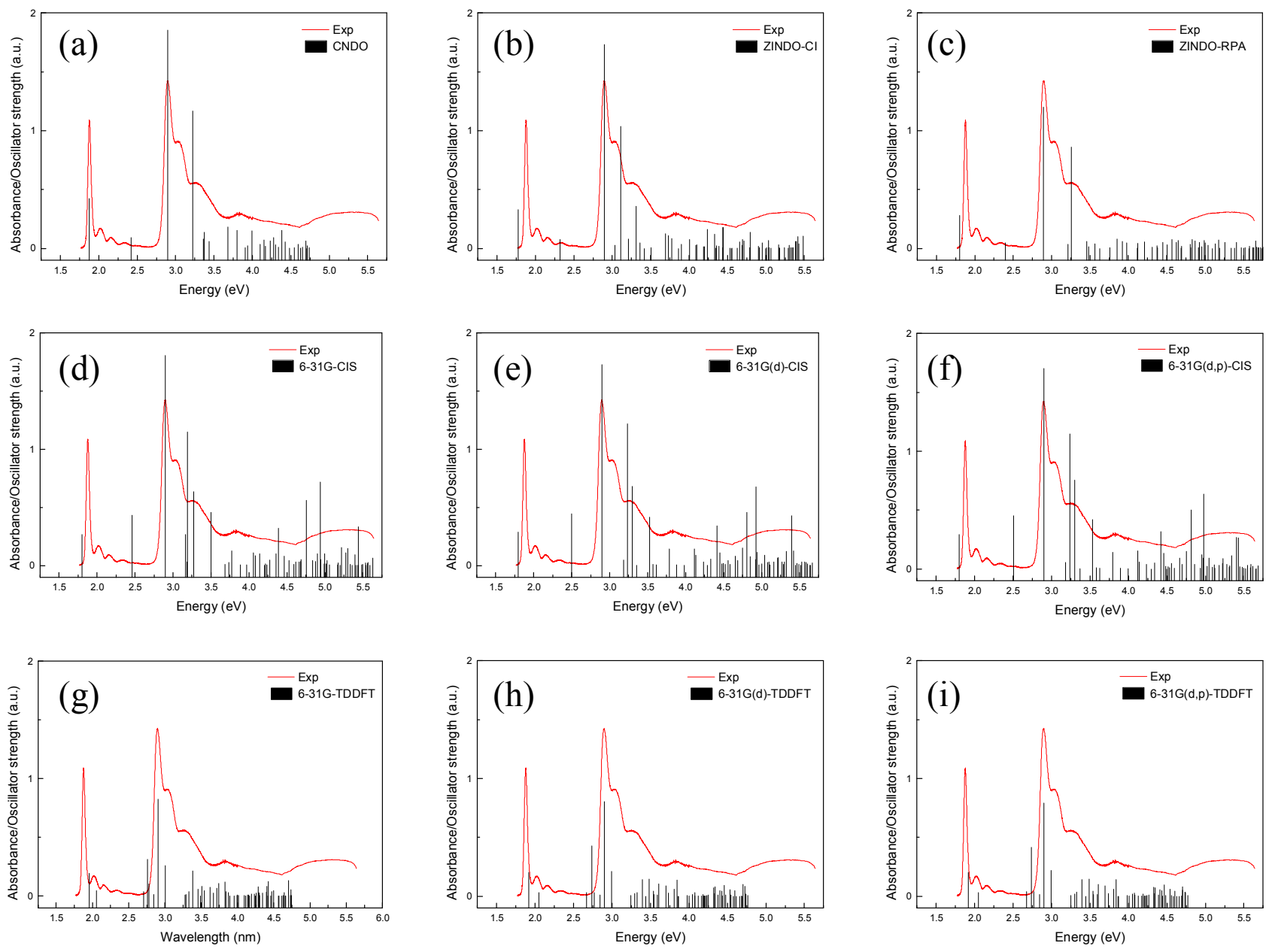

**Figure 7** Calculated UV-Vis absorption spectra of chlorophyll *a* molecule at different level of theories and basis sets: (a)CNDO/s, (b)ZINDO-CIS, (c)ZINDO-RPA, (d)RHF/CIS/6-31G, (e)RHF/CIS/6-31G(d), (f)RHF/CIS/6-31G(d,p), (g)TDDFT/6-31G, (g)TDDFT/6-31G(d), (g) TDDFT/6-31G(d,p). The continues curve in each graph represents UV-Vis absorption spectrum that observed experimentally[4].

The semi-empirical methods are still unmatched with any ab initio methods used in this study from the point of view of accuracy. However, as discussed above, the lack of fine features that might be important for certain purposes becomes their weak point. Here, the TDDFT methods give the optimum performance among other methods. As listed in Table 1, the best TDDFT procedure used in this study, TDDFT/6-31G(d,p) has the correction factor in the same level of CNDO/s but provides more complete picture of the electronic transitions. This is reflected by the absorption fine features in Figure 7(i) with respect to Figure 7(a). If we compare TDDFT/6-31G(d,p) with better semi-empirical methods, such as ZINDO-CI or ZINDO-RPA, this method still provides more complete transitions with slightly lower correction factor up to 9 %. Instead of its lower correction factor, it provides a much more complete feature as presented in Figure 7(i) with respect to Figure 7(b) and (c).

**Table 1** Correction factor at different level of theories and basis sets.

| Method | Correction factor |
|---|---|
| CNDO/S | 0.8260 |
| ZINDO-CI | 0.8400 |
| ZINDO-RPA | 0.8980 |
| CIS/6-31G | 0.6245 |
| TDDFT/6-31G | 0.8170 |
| CIS/6-31G(d) | 0.6300 |
| TDDFT/6-31G(d) | 0.8225 |
| CIS/6-31G(d,p) | 0.6325 |
| TDDFT/6-31G(d,p) | 0.8245 |

## 4 Conclusions

In this work, the vertical electronic transitions of chlorophyll *a* were calculated using both semi-empirical and ab initio methods to provide both contributed molecular orbitals and UV-Vis absorption spectra. The SAAP (spin-adapted antisymmetrized product) treatment on both RHF/CIS and TDDFT calculations reveals that the main molecular orbitals contributed in the electronic transitions are HOMO-1, HOMO, LUMO and LUMO+1 independent of level of theory and basis sets. In this study, both the addition of polarization functions and DFT treatment in molecular wavefunctions increase the accuracy of calculated UV-Vis absorption spectra. This concluded due to the increase of the long range MOs interactions spatially. Here, the effect of DFT treatment is more pronounced to their delocalized nature in their wavefunctions. The effect of the addition of polarization functions is more obvious in RHF/CIS treatment due their localized wavefunction. TDDFT/6-31G(d,p), the procedure with the highest level of theory in this study, provides more complete transitions with slightly lower correction factor up to 9 % with respect to ZINDO-RPA.

## Acknowledgements

This work was supported by ITB Research Grant Project 2009 with contract number 270/K01.7/PL/2009. The authors would like to thank to Alex A. Granovsky, Department of Chemistry, Moscow State University and M. A. Martoprawiro, Department of Chemistry, ITB for various discussions and suggestions about the ab initio calculation of electronic transitions spectra.

## References

[1] Barazzouk, S., Hotchandani, S., *Enhanced charge separation in chlorophyll a solar cell by gold nanoparticle*, J. Appl. Phys. **96**, 7744-7746, 2004.

[2] Shimatani, K., Tajima, H., Komino, T., Ikeda, S., Matsuda, M., Ando, Y., Akiyama, H., *The Electroluminescence Spectrum of Chlorophyll a*, Chem. Lett. **34**, 948-949, 2005.

[3] Linnanto, J., Korppi-Tommola, J., *Spectroscopic properties of Mg-chlorin, Mg-porphin and chlorophylls a, b, c1, c2, c3 and d studied by semi-empirical and ab initio MO/CI methods*, Phys. Chem. Chem. Phys. **2**, 4962-4970, 2000.

[4] Du, H., Fuh, R.-C., Li, J.Z., Corkan, L.A., Lindsey, J.S., *PhotochemCAD HD 1.1*, Carnegie Mellon University and North Carolina State University, 1998.

[5] Vokacova, Z., Burda, J. V., *Computational study on the spectral properties of the selected pigments from Various Photosystems: Structure-Transition energy relationship*, J. Phys. Chem. A **111**, 5864-5878, 2007.

[6] Hasegawa, J., Ozeki, Y., Ohkawa, K., Nakatsuji, H., *Theoretical Study of the Excited States of Chlorin, Bacteriochlorin, Pheophytin a, and Chlorophyll a by the SAC/SAC-CI Method*, J. Phys. Chem. B **102** 1320-1326, 1998.

[7] Parusel, A., Grime, S., *DFT/MRCI calculations on the excited states of porphyrin, hydroporphyrins, tetrazaporphyrins and metalloporphyrins*, *J. Porphyrins Phthalocyanines* **5**, 225-232, 2001.

[8] Dreuw, A., Head-Gordon, M., *Single-Reference ab Initio Methods for the Calculation of Excited States of Large Molecules*, Chem. Rev. **105**, 4009-4037, 2005.

[9] Staemmler, V., *Introduction to Hartree-Fock and CI Methods*, In J. Grotendorst, S. Blügel, D. Marx (Eds.), *Computational Nanoscience: Do It Yourself!*, John von Neumann Institute for Computing, Jülich, NIC Series, **31**, 1-18, 2006).

[10] Rohringer, N., Peter, S., Burgdörfer, J., *Calculating state-to-state transition probabilities within time-dependent density-functional theory*, Phys. Rev. A **74**, 042512-1-7, 2006.

[11] Doltsinis, N.L., *Time-Dependent Density Functional Theory*, In J. Grotendorst, S. Blügel, D. Marx (Eds.), *Computational Nanoscience: Do It Yourself!*, John von Neumann Institute for Computing, Jülich, NIC Series, **31**, 357-373, 2006.

[12] Granovsky, A.A., PC GAMESS/Firefly version 7.1.F, www http://classic.chem.msu.su/gran/gamess/index.html.

[13] DelBene, J., Jaffe, H.H., *Use of the CNDO Method in Spectroscopy. I. Benzene, Pyridine, and the Diazines*, J. Chem. Phys. **48**, 1807-1813, 1968.

[14] DelBene, J., Jaffe, H.H., *Use of the CNDO Method in Spectroscopy. II. Five ‐ Membered Rings*, J. Chem. Phys. **48**, 4050-4055, 1968.

[15] J. Ridley, M. C. Zerner, *An Intermediate Neglect of Differential Overlap Technique for Spectroscopy: Pyrrole and the Azines*, Theor. Chim. Acta **32**, 111-134, 1973.

[16] Thompson, M., Zerner, M.C., *A theoretical examination of the electronic structure and spectroscopy of the photosynthetic reaction center from Rhodopseudomonas viridis*, J. Am. Chem. Soc. **113**, 8210-8215, 1991.

[17] Bacon, A.D., Zerner, M.C., *An intermediate neglect of differential overlap theory for transition metal complexes: Fe, Co and Cu chlorides*, *Theor. Chim. Acta* **53**, 21-54, 1979.